\documentclass[a4paper,11pt]{article}
\pdfoutput=1 

\usepackage{jcappub} 

\usepackage[T1]{fontenc}
\usepackage{amssymb}
\usepackage{float}

\title{\boldmath Non-singular and Cyclic Universe from the Modified GUP}

\author[a]{Maha Salah,}
\author[b]{Fay\c{c}al Hammad,}
\author[c]{Mir Faizal}
\author[d]{and Ahmed Farag Ali}

\affiliation[a]{Department of Mathematics, Faculty of Science, Cairo University, Giza, 12613, Egypt.}
\affiliation[b]{Physics Department \& STAR Research Cluster, Bishop's University,\\
2600 College Street, Sherbrooke (QC), J1M 1Z7, Canada.\\Physics Department, Champlain College-Lennoxville,\\
2850 College Street, Sherbrooke (QC), J1M 0C8, Canada.}
\affiliation[c]{Department of Physics and Astronomy, University of Lethbridge,\\Lethbridge,  Alberta, T1K 3M4, Canada.}
\affiliation[d]{Department of Physics, Faculty of Science, Benha University, Benha, 13518, Egypt.}
\emailAdd{abdelmoneim.maha@gmail.com}
\emailAdd{fhammad@ubishops.ca}
\emailAdd{mirfaizalmir@gmail.com}
\emailAdd{ahmed.ali@fsc.bu.edu.eg}

\abstract{In this paper, we investigate the  effects of a new version of the generalized uncertainty
principle (modified GUP) on the dynamics of the Universe. As the modified GUP will modify the
relation between the entropy and area of the apparent horizon, it will also deform the Friedmann equations within Jacobson's
approach. We explicitly find these deformed Friedmann equations governing the modified GUP-corrected dynamics of
such a Universe. It is shown that the modified GUP-deformed Jacobson's approach implies an upper bound for
the density of such a Universe. The Big Bang singularity can therefore also be avoided using the modified GUP-corrections to horizons' thermodynamics. In fact, we are able to analyze the pre Big Bang state of the Universe. Furthermore, the equations imply that the expansion of the Universe will come to a halt and then will immediately be followed by a contracting phase. When the equations are extrapolated beyond the maximum rate of contraction, a cyclic Universe scenario emerges.}

\begin{document}
\maketitle
\flushbottom

\section{Introduction}\label{sec:1}
One of the important results in the search for quantum gravity has been the emergence of the concept of
minimum length. After the universal acceptance of the Planck length $l_{P}$ as the lower bound on any physical scale,
first introduced in Ref.~\cite{Mead}, several approaches towards understanding physics at this scale, based either on
string theory or other quantum gravity paradigm \cite{Veneziano, Amati, Gross, Yonega, Konishi, Guida, Maggiore1,
Maggiore2, Maggiore3, Garay, Kempf1, Kempf2, Brau, Scardigli, Hossenfelder1, Bambi, Ali1,Ali:2011ap,Ali:2011fa,Das:2010zf,Ali:2012mt}, have suggested that
there should be a minimum length in Nature because Heisenberg's uncertainty principle might actually be generalized
in such a way that a fundamental uncertainty on position is increased by new momentum-dependent terms.

In the generalized uncertainty principle (GUP), one finds that the product of the uncertainty on position
$\Delta x$ and the uncertainty on momentum $\Delta p$ is a function of the uncertainty $\Delta p$ (see , e.g.
the recent review \cite{Hossenfelder}). This implies a nontrivial increase in uncertainty on position with
respect to the usual uncertainties of quantum mechanics as one increases the energy of the probe.
The implications of this concept of GUP have extensively been investigated in recent literature, especially
in the domain of black hole thermodynamics as well as in cosmology. In the former, it has been shown that the
GUP is responsible for bringing corrections to the Bekenstein-Hawking entropy formula
\cite{Medved, Ali2, Awad, Bina}.
The resulting   entropy on the horizon acquires logarithmic deviations with
respect to the pure area-law.
However, the Bekenstein-Hawking entropy formula also holds for the apparent horizon of the Universe.
Hence, it is possible to consider similar modifications to the Bekenstein-Hawking entropy formula for the apparent horizon of the Universe,
and this can have application in cosmology. In fact, in
 the domain of cosmology \cite{Akbar, Majumder, Lidsey, Awad}, the authors
of Ref.~\cite{Awad} have also demonstrated  that if one applies this corrected entropy formula to the apparent
cosmological horizon of the Universe, an upper bound on the density of the Universe emerges and the Big Bang
singularity dissolves. In addition, and still in the domain of cosmology, the authors of Ref.~\cite{Jalalzadeh} have successfully applied
the GUP to obtain a novel relation between entropy, the apparent horizon and the cosmological holographic principle.

As it happens, though, there are two versions of the GUP. In the simplest version of the GUP, the product
of the uncertainty $\Delta x$ and the uncertainty $\Delta p$
acquires a single additional term compared to the usual Heisenberg uncertainty principle. This
additional term is quadratic in $\Delta p$.
It is this additional term that is responsible for
the emergence of the maximum density and for avoiding the Big Bang singularity~\cite{Ali2}.
In the second version of the GUP \cite{Ali1},
which we call here the modified GUP, one finds, in addition to the quadratic term in $\Delta p$ on the
right-hand side of the inequality, another term linear in the uncertainty $\Delta p$.
The modified GUP is consistent with the existence of minimum length as well as doubly special relativity \cite{DSR}. This makes it also consistent with a large number of approaches to quantum gravity, such as discrete space-time \cite{17},  spontaneous symmetry breaking of
Lorentz invariance in string field theory \cite{18}, ghost condensation \cite{18aa}, space-time foam models \cite{19}, spin-networks in loop quantum gravity \cite{20}, non-commutative geometry \cite{21, 21aa},  and Horava-Lifshitz gravity \cite{22}.
As modified GUP is consistent with such a large number of theories,
it is important to analyze the consequences of modified GUP for cosmology by considering its effects on the dynamics of the Universe. The aim of this paper is to investigate and study such effects.

The outline of the present paper is as follows. In Sec.~\ref{sec:2}, we review the derivation of Friedmann equations from the thermodynamics approach. In Sec.~\ref{sec:3}, we briefly recall the modified formulation of the GUP and then use it to calculate the entropy that might be associated to horizons. In Sec.~\ref{sec:4}, we apply the formula thus found for entropy to the case of an apparent horizon
of the Universe and deduce the Friedmann equations as well as the density of the Universe using the general formula introduced in Sec.~\ref{sec:2}. We then solve the equations, and analyze and discuss the physical meaning of the solution. We end this
paper with a brief conclusion section.

\section{Friedmann Equation  in the  Thermodynamic Approach}\label{sec:2}
In this section, we shall briefly review the main steps in the derivation of Friedmann equations based on
the thermodynamic approach\cite{Akbar, Awad, cai}.  The starting point is the metric of the $4$-dimensional
Friedmann-Lema\^{\i}tre-Robertson-Walker (FLRW) Universe, which can be written in the form
\begin{eqnarray}
ds^2=-\mathrm{d}t^{2}+a^{2}(t)\left(\frac{\mathrm{d}r^{2}}{1-kr^2}+r^2d\Omega^2\right), \label{metric}
\end{eqnarray}
where $a(t)$ is the time-dependent positive scale factor and $\mathrm{d}\Omega^2=\mathrm{d}\theta^{2}+\sin^{2}\theta\mathrm{d}\phi^{2}$ is the line element on the unit two-sphere. The value of $k$ depends on the geometry of the Universe; the value $k =0$ corresponds to a flat Universe, $k=1$ is for a closed Universe, and $k=-1$ is for an  open Universe.  Now, the radius $r_{A}(t)$ of the apparent horizon corresponding to the metric (\ref{metric}) is found from the condition $\nabla_{\mu}R\nabla^{\mu}R=0$, where $R(t,r)=a(t)r$ is the areal radius of the two-sphere. On then obtains,
\begin{eqnarray}
\label{app-radius}
r_A(t)= \frac{1}{\sqrt{H^2+k/a^2}},
\end{eqnarray}
where $H=\dot{a}/a$ is the Hubble parameter; the dot standing for a time-derivative.
If it is assumed that the matter in  the FLRW Universe forms a perfect fluid with the four-velocity $u^{\mu}$,  then the energy-momentum tensor can be written as
\begin{eqnarray}\label{Energy-Momentum}
T_{\mu\nu}=(\rho+p)u_{\mu}u_{\nu}+pg_{\mu\nu},
\end{eqnarray}
where $\rho$ is the energy density of the perfect fluid and  $p$ is its pressure.
The conservation equation written in terms of the energy-momentum tensor, $\nabla^{\mu}T_{\mu\nu}=0$, can then be used to obtain,
\begin{eqnarray}
\dot{\rho}+3H(\rho+p)=0. \label{Continuity}
\end{eqnarray}
On the other hand, the  Misner-Sharp energy, $E=\rho V$, corresponding to the total matter present within the apparent horizon of the Universe, whose apparent volume is $V=\frac{4}{3}\pi r^{3}_{A}$\footnote{Note that the spatial curvature $k$ might take on three possible values and that only for the case $k=0$ does the proper radius $a(t)\int dr/\sqrt{1-kr^2}$ coincide with the co-moving radius $a(t)r$ of the sphere enclosing the volume of interest. This method works for all three cases of curvature because the radius of interest here is not the proper radius but the apparent horizon radius $r_{A}$, given by (\ref{app-radius}), valid for $k=0,\pm1$, and on which the usual definition of the volume of a sphere can be used. In other words, we are using the apparent three dimensional sphere.}, simply reads, $E=\frac{4}{3}\pi\rho r^{3}_{A}$. Therefore, when differentiating this expression, we find the infinitesimal change in the total energy of the perfect fluid during an infinitesimal interval of time $\mathrm{d}t$:
\begin{eqnarray}
\mathrm{d}E=4\pi\rho r^{2}_{A}\mathrm{d}r_{A}-4\pi r_{A}^3(\rho+p)H\mathrm{d}t. \label{denergy1}
\end{eqnarray}
To obtain this form, use have been made of the continuity equation (\ref{Continuity}) to express the differential $\mathrm{d}\rho$ in terms of the infinitesimal time-interval $\mathrm{d}t$.

Next, a work density $W$, to be associated with the perfect fluid, is introduced. This work density is extracted from the energy-momentum tensor of the perfect fluid by projecting (\ref{Energy-Momentum}) onto the normal direction to the apparent horizon, i.e. $W=-\frac{1}{2}h^{\mu\nu}T_{\mu\nu}$, where $h_{\mu\nu}$ is the two-metric on the normal direction. This work density is found to be \cite{Awad, Akbar,Hayward:1997jp}:
\begin{eqnarray}\label{work}
 W=\frac{1}{2}\left(\rho-p\right).
\end{eqnarray}
Let us now substitute this work density, together with the total differential (\ref{denergy1}), inside the first law of thermodynamics, $\mathrm{d}E=\delta Q+ W\mathrm{d}V$, where $\delta Q$ is the change of the total heat associated with the perfect fluid\footnote{Note that the first law of thermodynamics as we applied here is, as rightly pointed out by the anonymous referee, in principle incomplete. Indeed, we considered here only the work density (\ref{work}) coming from the energy-momentum tensor of the perfect fluid without including the chemical potential contribution $\mu dN$ when writing the first law of thermodynamics. The first reason is that, although we maintain the dark component of the Universe in this paper, we do not include any unknown non-zero chemical potential that would be attributed to dark energy (see \textit{e.g.} Ref.~\cite{Lima} for a discussion on this issue). On the other hand, since photons are much more abundant than baryons and leptons in the Universe, we rely on the usually good cosmological approximation that discards the $\mu dN$ term from the first law.}. According to the second principle of thermodynamics for reversible processes, which we assume to apply for the perfect fluid as the Universe evolves, we also have $\delta Q=T\mathrm{d}S$, where $T$ is the temperature and $\mathrm{d}S$ is the corresponding variation of entropy. The first law then yields,
\begin{eqnarray}\label{FirstTdS}
T\mathrm{d}S&=&2\pi(\rho+p)r^{2}_{A}\mathrm{d}r_{A}-4\pi r_{A}^3(\rho+p)H\mathrm{d}t\nonumber\\
&=&4\pi(\rho+p)\left(\frac{\dot{r}_{A}}{2Hr_{A}}-1\right)Hr^{3}_{A}\mathrm{d}t.
\end{eqnarray}

Now, since the radius $r_{A}$ of the apparent horizon is already given in terms of the Hubble parameter $H$ via identity (\ref{app-radius}), it is clear that in order to arrive at a Friedmann-like equation, we need only find the temperature $T$, as well as the entropy $S$, to be associated with the apparent horizon of the Universe. The first thing that comes to mind is of course the Hawking temperature associated to black holes' event horizons, as well as the Bekenstein-Hawking entropy of that kind of horizons. What we have here, however, is not a black hole's event horizon, but a cosmological apparent horizon. Therefore, although it seems natural to apply the formalism of black hole thermodynamics to the whole Universe, one still needs a physical justification for such a leap.

By now, it has been widely argued in the literature (see, e.g. Ref.~\cite{Bak} for an earlier one and Ref.~\cite{Viaggiu} for a more recent one, as well as the references therein) that it is physically justified to apply the thermodynamics of event horizons to the case of apparent horizons based on the holographic principle \cite{tHooft,Susskind,Fischler}. Indeed, although the Hawking temperature was originally found by studying a scalar field using quantum field theory techniques in the near-horizon curved spacetime, the fact that the entropy found for black holes is encoded on their boundary rather than their bulk, has given rise to the idea that physical degrees of freedom are always located on the boundaries. This holographic behavior of degrees of freedom has subsequently been suggested as a simple alternative for dark energy when it comes to explaining the accelerated expansion of the whole Universe \cite{Easson1,Easson2,Komatsu} (see also \cite{hammad}). The mere fact that this idea has allowed to recover Friedmann equations is actually another argument in favor of the whole approach. In this paper, we shall therefore adopt this widely held view and adapt the thermodynamics formalism of black holes' event horizons to the apparent cosmological horizon.

Before proceeding further, we would like to stress here the fact that our approach is based on the application of the first law of
thermodynamics, as well as the concept of Hawking's temperature, to the Universe's apparent horizon just as it was done in Ref.~\cite{Awad}.
The apparent cosmological horizon is chosen over both the event and Hubble horizons for the reason that these latter do not always exist, unlike the apparent horizon which exists for all values of $k$ and to which the two others reduce in the case of a flat Universe ($k=0$).
Indeed, it has already been pointed out in detail in Ref.~\cite{Wang} that one should be careful and distinguish between apparent and cosmological event horizons when studying the thermodynamics of the Universe, as the two categories of horizons coincide generally only for the case of a de Sitter space-time.

In addition, it has been shown in Ref.~\cite{Cai2} that the FLRW apparent horizon, being defined as the marginally trapped surface with vanishing expansion, is endowed with a Hawking-like temperature related to its surface gravity. This is due to the fact that, whereas for stationary black holes one has a time-like Killing vector field from which one can define a conserved energy for a particle moving in the space-time, for the case of an apparent horizon one does not have a time-like Killing field but a Kodama vector field $K^{\mu}$. This vector field satisfies $K_{\mu}K^{\mu}=R^{2}/r^{2}_{A}-1$ and, hence, becomes time-like only inside the apparent horizon where the areal radius satisfies, $R<r_{A}$. One can then also define a conserved energy for a particle moving inside the apparent horizon and, in analogy to what is done for the case of black holes' horizons \cite{Parikh}, one can compute the tunneling amplitude across the apparent horizon. The corresponding temperature found is indeed the Hawking temperature \cite{Cai2}.

Furthermore, the approach based on the apparent horizon allows one to recover the Friedmann equation, as we shall see below, for all three values of $k$. This stems from the fact that in the application of the holographic principle, the specific spatial geometry of the Universe does not matter, as long as spherical symmetry of the boundary holds. This fact is actually yet another reason to apply black holes' thermodynamics to apparent horizons regardless of the value of $k$.

Let us therefore adopt the Hawking temperature for the apparent horizon by associating to the latter, just as it is usually done for the case of black holes, the temperature $T=\kappa/2\pi$, where, this time, $\kappa$ is the surface gravity evaluated on the apparent horizon. One finds \cite{cai},
\begin{eqnarray}\label{kappa}
\kappa&=&\frac{1}{2\sqrt{-h}}\partial_a(\sqrt{-h}h^{ab}\partial_{b}R)\nonumber\\
&=&-\frac{1}{r_{A}}\left(1-\frac{\dot{r}_A}{2Hr_A}\right),
\end{eqnarray}
where $h_{ab}$, appearing in the first line, is again the two-metric of the $(t,r)$-space in (\ref{metric}).

As for entropy, one also adopts the Bekenstein-Hawking area-law and associates to the apparent horizon an entropy $S=A/4G$, where $A=4\pi r^{2}_{A}$ is the area of the horizon whose radius is $r_{A}$ and $G$ is Newton's gravitational constant. Therefore, using both (\ref{kappa}) and this linear area-law for entropy, it is possible to express $T\mathrm{d}S$ appearing in the left-hand side of (\ref{FirstTdS}) in terms of the Hubble parameter and the radius $r_{A}$:
\begin{eqnarray}
T\mathrm{d}S=-\frac{1}{G}\left(1-\frac{\dot{r}_A}{2Hr_A}\right)\mathrm{d}r_{A}. \label{TdS}
\end{eqnarray}
Substituting (\ref{TdS}) into the left-hand side of (\ref{FirstTdS}), after using the fact, as it follows from (\ref{app-radius}), that $\mathrm{d}r_A=-Hr_A^3\left(\dot{H}-k/a^2\right)\mathrm{d}t$, one easily recovers the dynamical Friedmann equation \cite{cai},
\begin{eqnarray}
\dot{H}-\frac{k}{a^2}=-4\pi G(\rho+p).
\label{dotH}
\end{eqnarray}
Then, after using the continuity equation (\ref{Continuity}), this first-order differential equation can easily be integrated to yield the well-known general Friedmann equation for an FLRW Universe:
\begin{eqnarray}\label{Friedmann}
H^2+\frac{k}{a^2}=\frac{8\pi G}{3}\rho.
\end{eqnarray}

Note that, as already emphasized above, the Friedmann equation thus obtained is valid for all three possible values of $k$. This couldn't have been found had we used the event or Hubble horizon instead of the apparent horizon.

Now, this derivation, as we see from (\ref{TdS}), critically depended on the use of the linear entropy-area law, $S = A/4G$, for the apparent horizon. Therefore, any modification of this relation between the entropy of the apparent horizon and its area will modify the Friedmann equation for an FLRW Universe. Let us therefore examine here the consequences of such a modification by assuming the following general form of a modified entropy-area law for the apparent horizon:
\begin{eqnarray}\label{GeneralEntropy}
S=\frac{f(A)}{4G},
\end{eqnarray}
where $f(A)$ is any arbitrary smooth function of the area $A$ of the apparent horizon. Taking the differential of both sides of this identity, we find
\begin{eqnarray}\label{dS}
\mathrm{d}S=\frac{f^{\prime}(A)}{4G}\mathrm{d}A=\frac{f^{\prime}(A)}{G}2\pi r_{A}\mathrm{d}r_{A}.
\end{eqnarray}
where the prime stands for a derivative with respect to the area $A$. Therefore, instead of identity (\ref{TdS}), we will have to use the following identity:
\begin{eqnarray}
T\mathrm{d}S=-\frac{f'(A)}{G}\left(1-\frac{\dot{r}_A}{2Hr_A}\right)\mathrm{d}r_{A}. \label{ModifiedTdS}
\end{eqnarray}
Then, by substituting this in the left-hand side of (\ref{FirstTdS}), we find instead of the dynamical Friedmann equation (\ref{dotH}), the following differential equation
\footnote{Note that the use of identity (\ref{FirstTdS}) in (\ref{ModifiedTdS}) is not inconsistent with the fact that here we apply it for the general case of extended theories of gravity that yield a general area-law $f(A)$ in (\ref{GeneralEntropy}). Indeed, identity (\ref{FirstTdS}) does not rely on general relativity to hold but on the general laws of thermodynamics. In fact, as has been shown in Ref.~\cite{Jacobson1}, Einstein equations themselves are derivable from such laws. Moreover, the main point of Ref.~\cite{cai} was to obtain Friedmann equations from thermodynamics alone as we saw from the derivation of Eq. (\ref{dotH}). On the other hand, the apparent horizon used in (\ref{FirstTdS}), and extracted from the FLRW geometry, does not rely on general relativity either as it is obtained from purely geometric arguments in (\ref{app-radius}). Furthermore, one uses FLRW geometry as a background spacetime on which one tests one's dynamical equations. The use of FLRW geometry does not require general relativity. Such a background has been used in the literature to reconstruct models of $f(R)$-gravity, see e.g. Ref.~\cite{Carloni}.}:
\begin{eqnarray}
\left(\dot{H}-\frac{k}{a^2}\right)f^{\prime}(A)= -4\pi G\left(\rho+p\right) \label{FR1}.
\end{eqnarray}
Finally, by expressing the left-hand side in terms of $\mathrm{d}r_{A}$ thanks again to the fact that $\mathrm{d}r_A=-Hr_A^3\left(\dot{H}-k/a^2\right)\mathrm{d}t$, and using on the right-hand side the continuity equation (\ref{Continuity}), the above differential equation transforms into,
\begin{eqnarray}
f'(A)\frac{\mathrm{d}r_{A}}{r^{3}_{A}}=-\frac{4\pi G}{3}\mathrm{d}\rho.
\end{eqnarray}
which integrates to yield the modified Friedmann equation \cite{Awad}:
\begin{eqnarray}
\frac{2G}{3}\rho=-\int f^{\prime}(A)\frac{\mathrm{d}A}{A^{2}}\label{FR2}.
\end{eqnarray}
Note that this formula is what the general formula, given in Ref. \cite{Awad}, reduces to when setting $n=3$ there. This formula says that for each different area-law of the entropy one happens to associate to the apparent horizon one gets a different modified Friedmann equation.

Now, as this approach is based on the assumption that black hole thermodynamics might be adapted to the study of the apparent horizon of the Universe, it is natural to also assume, in accordance with the holographic principle, that any modification of the areal-law coming from the physics of black holes would imply a similar modification to the area-law that one should use to study the apparent horizon of the Universe.
In the next two sections, we will find that the new relation between entropy and the horizon area, as it is derived in the literature from the generalized uncertainty principle applied to the case of black holes, will yield yet another modified Friedmann equation for the Universe when adapted to the case of the apparent horizon.

\section{The Modified GUP and Apparent Horizon Entropy}\label{sec:3}
The generalized uncertainty principle, or GUP for short, that most of the approaches to quantum gravity seem to agree upon
(see, e.g. Ref.~\cite{Hossenfelder}) is a generalization of Heisenberg's uncertainty principle given by the inequality
$\Delta x\Delta p\geq1/2$ to an inequality of the form $\Delta x\Delta p\geq1/2(1+\alpha^{2}l_{P}^{2}\Delta p^{2})$ where
$\alpha$ is a dimensionless numerical factor that depends on the model used to investigate the physics at Planck lengths
\footnote{We shall use throughout this paper unites in which $\hbar=c=1$.} and $l_{P}$ is the Planck length.
In the new, or modified, version of the GUP, one finds, in addition to the quadratic term $\Delta p^{2}$, a linear term in $\Delta p$
as follows \cite{Ali1,Ali:2011ap,Ali:2011fa,Das:2010zf,Ali:2012mt}:
\begin{equation}\label{1}
\Delta x\Delta p\geq\frac{1}{2}\left[1-\frac{4\sqrt{\mu}}{3}\alpha l_{P}\Delta p+2(1+\mu)\alpha^{2}l^{2}_{P}\Delta p^{2}\right].
\end{equation}
Whereas the factor $\alpha$ in the purely quadratic version of the GUP is left unspecified, the factor $\alpha$ in the version (\ref{1})
above has an upper bound whose exact value is left to be determined experimentally \cite{Ali:2011fa}. $\mu$ in inequality (\ref{1}) is another
dimensionless factor. The value of this factor is of order unity and depends on the quantum gravity model used to extract inequality (\ref{1}) \footnote{In Ref.~\cite{Ali:2012mt} this value has been set at $(2.82/\pi)^2$ by assuming that a black hole's particle emission peaks in momentum $\langle p\rangle$ according to Wien's law $\langle p\rangle\sim2.82\,T_{H}$, where $T_{H}$ is the Hawking temperature of the black hole. Then, since Hawking's temperature is related to the uncertainty $\Delta p$ of the emitted particles as $T_{H}=\Delta p/\pi$ \cite{adler} and if one assumes that $\langle p\rangle=\sqrt{\mu}\Delta p$, for some positive factor $\mu$ \cite{Ali:2012mt}, the estimate $\mu=(2.82/\pi)^2$ follows. As we shall see later, however, as long as this parameter remains of the order unity, its exact numerical value does not matter as it does not affect our conclusions.}. Let us now follow,
using this modified version of the GUP, the usual procedure for extracting the horizon's entropy based on the standard GUP \cite{adler,Ali:2013qza}.

By extracting the uncertainty on momentum $\Delta p$ in terms of the uncertainty on position $\Delta x$ from the above inequality, one finds
\begin{equation}\label{2}
\Delta p\geq\frac{\Delta x}{\gamma^{2}}+\frac{2}{3\gamma}\sqrt{\frac{\mu}{2(1+\mu)}}-
\sqrt{\left(\frac{\Delta x}{\gamma^{2}}+\frac{2}{3\gamma}\sqrt{\frac{\mu}{2(1+\mu)}}\right)^{2}-\frac{1}{\gamma^{2}}},
\end{equation}
where we have set, for ease of notations, $\gamma^{2}=2(1+\mu)\alpha^{2}l^{2}_{P}$ and chose
the minus sign outside the second square-root in order to recover Heisenberg's uncertainty principle
in the limit $l^{2}_{P}\rightarrow 0$. Now we shall follow the usual steps that lead to the minimum
increase in the event horizon's area $A$ \cite{adler,Ali:2013qza}. First, trading the energy $E$ for the uncertainty $\Delta p$
on momentum in the above expression \footnote{We should note here that in performing this step we have combined the usual relativistic dispersion relation between momentum and energy with the GUP. One might wonder, however, why GUP would not imply modifications to the dispersion relation, i.e, the so-called MDR \cite{20,MDR,Amelino}. The reason is that the GUP and MDR are two sides of the same coin. In practice, one either uses the heads or the tails to find the same result, but never both at the same time. Indeed, thanks to the identity, $\Delta x\Delta E\sim|{[x,p]}|\partial E/\partial p$, one is able to obtain a GUP in the form, $\Delta x\Delta E\sim 1+2\alpha p+...$, from an MDR in the form, $E=p+\alpha p^2+...$, by imposing the usual Heisenberg's uncertainty principle in commutator form, $[x,p]=i$. But one can also invert identity, $\Delta x\Delta E\sim|{[x,p]}|\partial E/\partial p$, to obtain MDR in the generalized commutator form, $|[x,p]|=1+2\alpha p+...$, by using the GUP and imposing the usual relativistic dispersion relation $E\sim p$. A good reference discussing the details of this subtlety can be found in, e.g., Refs.~\cite{Xiang,Amelino}.} allows one to find the lower bound on energy $E$ in terms of the uncertainty on
position $\Delta x$. Then, multiplying the resulting inequality by $\Delta x$ on both sides and using that the minimum increase
$\Delta A_{min}$ in area of the event horizon is related to the energy $E$ and spatial extension $\Delta x$ of the particle by
$\Delta A_{min}\geq 8\pi l^{2}_{P}E\Delta x$ \cite{Camelia}, leads to
\begin{equation}\label{3}
\Delta A_{min}\geq8\pi l^{2}_{P}\Delta x\left[\frac{\Delta x}{\gamma^{2}}+\frac{2}{3\gamma}\sqrt{\frac{\mu}{2(1+\mu)}}-
\sqrt{\left(\frac{\Delta x}{\gamma^{2}}+\frac{2}{3\gamma}\sqrt{\frac{\mu}{2(1+\mu)}}\right)^{2}-\frac{1}{\gamma^{2}}}\right].
\end{equation}
Finally, by using the assumption that entropy always increases, according to information theory and the holographic principle,
by a constant unit usually denoted $b$, whenever the horizon area $A$ increases by the elementary amount $\Delta A_{min}$, i.e.
$\mathrm{d}S/\mathrm{d}A=b/\Delta A_{min}$, we find the following differential equation for entropy:
\begin{equation}\label{4}
\frac{\mathrm{d}S}{\mathrm{d}A}=\frac{b}{8\pi\epsilon l^{2}_{P}}\left[\frac{A}{\beta}+\eta\sqrt{A}
-\sqrt{\left(\frac{A}{\beta}+\eta\sqrt{A}\right)^{2}-\frac{A}{\beta}}\right]^{-1},
\end{equation}
where we have set
\begin{eqnarray}\label{5}
\beta&=&\pi\gamma^{2},\nonumber
\\\eta&=&\frac{2}{3\gamma}\sqrt{\frac{\mu}{2\pi(1+\mu)}}.
\end{eqnarray}
The factor $\epsilon$ in (\ref{4}) has been introduced to fix the minimum value of the minimum increase $\Delta A_{min}$
and substituted $\Delta x$ by $\sqrt{A/\pi}$. The original argument was made using a
quantum particle near the horizon of a black hole, and it was argued that it
has a natural `size' of the order of the Schwarzschild radius $r_{S}$, i.e.
$\Delta x=2r_{S}=\sqrt{A/\pi}$ \cite{adler,Ali:2013qza}. However, the Bekenstein-Hawking   entropy formula also holds for the apparent horizon
of the Universe. Furthermore, this argument also hold  any
spherical symmetrical horizon, so it will also hold for the apparent horizon of the Universe. Hence, we apply this argument directly to the
apparent horizon, and argue that the GUP will also modify the  entropy-area law of the apparent horizon.
We see that in order to recover the original entropy-area law of  the apparent horizon   in the limit
$l^{2}_{P}\rightarrow0$, we should chose $b/\epsilon=\pi$. Note that the value found in Ref. \cite{Awad} for this ratio
is twice the present value. This is due to the extra factor of $1/2$ present in inequality (\ref{1}) used here compared to inequality (4.1) used in Ref.~\cite{Awad}. In the next section, we will use this new entropy-area law for the apparent horizon to analyze the modification to the Friedmann equation.

\section{The Modified GUP and the Cyclic Universe}\label{sec:4}
In this section, we shall examine the implications of the differential equation (\ref{4}), that gives entropy in terms of the area of the event horizon, on the cosmic density $\rho$. For that purpose, we shall use Eq.~(\ref{FR2}) that relates the cosmic density $\rho$ to the area $A$ of the Universe's apparent horizon.

First, note that the differential equation (\ref{4}) is of the form $\mathrm{d}S/\mathrm{d}A=f'(A)/4G$, where $f(A)$ is some function of the area
$A$ such that its first derivative with respect to the latter is,
\begin{equation}\label{f'(A)}
f'(A)=\frac{1}{2}\left[\frac{A}{\beta}+\eta\sqrt{A}
-\sqrt{\left(\frac{A}{\beta}+\eta\sqrt{A}\right)^{2}-\frac{A}{\beta}}\right]^{-1},
\end{equation}
where we have used the previous choice of $b/\epsilon=\pi$ and the fact that in the natural units used here, $l_{P}^{2}=G$. Using the thermodynamic approach reviewed in Sec.~\ref{sec:2} and based on a different form of $f'(A)$ than expression (\ref{f'(A)}), the authors in Ref.\cite{Awad} have found a specific Friedmann equation. In using the above expression for $f'(A)$ we will find a different Friedmann equation, with different and interesting physical implications.

When substituting expression (\ref{f'(A)}) inside the general integral in the right-hand side of (\ref{FR2}), we find the following expression of the density $\rho$ in terms of the area $A$:
\begin{eqnarray}\label{7}
\frac{4G\rho}{3}&=&\frac{1}{A}+\frac{2\beta\eta}{3A^{3/2}}-\frac{2\xi^{3/2}}{3\beta(1-\beta\eta^2)A^{3/2}}
-\frac{\eta(\beta\eta^{2}-1+\eta\sqrt{A})\xi^{1/2}}{(1-\beta\eta^{2})^{2}A}\nonumber
\\&&-\frac{\eta}{\sqrt{\beta}(1-\beta\eta^{2})^{5/2}}\sin^{-1}\left[\frac{\sqrt{\beta}(\beta\eta^{2}-1)}{\sqrt{A}}+\eta\sqrt{\beta}\right]+C,
\end{eqnarray}
where we have set
\begin{eqnarray}\label{8}
\xi=A+2\beta\eta A^{1/2}+\beta^{2}\eta^{2}-\beta.
\end{eqnarray}
The constant of integration $C$ in (\ref{7}) can be fixed by noticing, as done in Ref.~\cite{Awad}, that for $A\rightarrow\infty$ we should find a cosmic density $\rho$ dominated by the cosmological constant $\Lambda$, i.e. that$\underset{\small{A \to \infty}}\lim\rho(A)=\Lambda$. With this requirement, (\ref{7}) leads to
\begin{equation}\label{9}
C=\frac{4G\Lambda}{3}+\frac{2+3\beta\eta^{2}}{3\beta(1-\beta\eta^{2})^{2}}+\frac{\eta\sin^{-1}(\eta\sqrt\beta)}{\sqrt{\beta}(1-\beta\eta^{2})^{5/2}}.
\end{equation}
Note that for $l_{P}\rightarrow0$, identity (\ref{7}) reduces, when using (\ref{9}), to $2G\rho/3=2G\Lambda/3+A^{-1}$. This, after expressing the area $A$ in terms of the radius of the apparent horizon, is actually nothing but the Friedmann equation displaying a cosmological constant, $H^{2}+k/a^{2}=\frac{8}{3}\pi G(\rho-\Lambda)$, whose consequences are discussed below.

We shall use identity (\ref{7}) to deduce below the upper bound on the cosmic density. To do so, however, we must first deduce the condition that the area $A$ should satisfy in order to make $\rho$ acquire a real value. Indeed, given the presence of the half-integer powers of $\xi$ in identity (\ref{7}), it is clear that the density $\rho$ might come out complex. So we must demand that $\xi$ be positive or null. In fact, this condition is also the same that avoids that the horizon's entropy, as constrained by (\ref{f'(A)}), comes out complex-valued. Indeed, for entropy $S$ to be real, the derivative $f'(A)$ as given by the right-hand side of (\ref{f'(A)}) should be real too. This, in turn, simply amounts to demand that the argument of the square root in (\ref{f'(A)}) be positive or null. This latter condition is in fact nothing but a requirement that $\xi\geq0$. Therefore, we have the following unique constraint that would make both the entropy $S$ and the density $\rho$ of the Universe real:
\begin{equation}\label{10}
A\geq\left(\sqrt{\beta}-\beta\eta\right)^{2}.
\end{equation}

Given that in an FLRW Universe whose scale factor is $a(t)$, the area of the apparent horizon $A$ is related to the Hubble parameter $H$ and the Gaussian curvature $k$ by $A=4\pi/(H^{2}+k/a^{2})$, substituting this inside identity (\ref{7}) gives the generalized Friedmann equation as implied by the modified GUP. More importantly, however, when substituting this value for $A$ in inequality (\ref{10}) the latter transforms into
\begin{equation}\label{11}
H^{2}+\frac{k}{a^{2}}\leq\frac{4\pi}{\left(\sqrt{\beta}-\beta\eta\right)^{2}}.
\end{equation}

Note that after substituting the definitions (\ref{5}) of $\beta$ and $\eta$ in this equation, the latter gives for the simple case of a flat Universe, $k=0$, the constraint $H^{2}\leq18/[\alpha^{2}l_{P}^{2}(3\sqrt{1+\mu}-\sqrt{2\mu})^{2}]$. This is a constraint on the early-times evolution of the Universe and, hence, on the inflationary energy scale. Using the fact that the numerical factor $\mu$ is of order unity, gives the following maximum value for the inflationary scale:
\begin{equation}
H^{2}_{max}\sim\frac{M_{P}^{2}}{\alpha^{2}},
\end{equation}
where $M_{P}$ is the Planck mass. For an upper bound of $10^{16}$ for $\alpha$ \cite{Ali:2011fa}, we recover the minimum value of $\sim1$ TeV for the the inflationary energy scale. Conversely, since the maximum energy scale for inflation should not exceed the Planck scale $M_{P}$, the above result constrains the minimum value that $\alpha$ could take on to be $\gtrsim 1$. As we shall see below, however, our conclusions do not depend on the upper and lower bounds of $\alpha$.

Now, Eq.~(\ref{11}) also suggests that the Hubble parameter can never exceed a given value and, hence, the Universe could be prevented from reaching a singularity. This fact becomes actually even more apparent after computing the maximum density $\rho_{max}$ as it follows from (\ref{7}) due to the lower bound imposed on the area $A$ by (\ref{10}). Indeed, notice first that when $A$ takes on the minimum value $A_{min}$ as given by the right-hand side of inequality (\ref{10}), $\xi$ in identity (\ref{7}) vanishes. Therefore, the latter identity yields
\begin{equation}\label{12}
\frac{4G}{3}\rho_{max}=\frac{3\sqrt{\beta}-\beta\eta}{3\left(\sqrt{\beta}-\beta\eta\right)^{3}}
+\frac{\pi\eta}{2\sqrt{\beta}(1-\beta\eta^{2})^{5/2}}+C,
\end{equation}
where we have chosen the value $-\pi/2$ for the inverse function $\sin^{-1}$ when its argument equals $-1$ instead of $(2m+1)\pi/2$ in order to keep the density $\rho_{max}$ positive. Thus, the result (\ref{12}) is the maximum value of the density that the Universe is allowed to reach if the modified GUP holds. The analysis done in Ref.~\cite{Awad} concerning the interpretation of the existence of this upper bound for the density of the Universe can be repeated here verbatim except for the intriguing presence of the inverse $\sin^{-1}$ function. This single difference brings actually non-trivial consequences as it will be apparent from the analysis to which we turn now.

In order to analyze the above result more rigorously, let us write the dynamical Friedmann equation by substituting (\ref{f'(A)}) inside the general formula (\ref{FR1}). This gives,
\begin{equation}\label{14}
\dot{H}-\frac{k}{a^{2}}=-8\pi G(\rho+p)\left[\frac{A}{\beta}+\eta\sqrt{A}
-\sqrt{\left(\frac{A}{\beta}+\eta\sqrt{A}\right)^{2}-\frac{A}{\beta}}\right],
\end{equation}
where of course one should substitute $4\pi/(H^{2}+k/a^{2})$ for $A$ in the above differential equation. Note that for $l_{P}\rightarrow0$, the content of the square brackets reduces to $1/2$ in this last identity and we recover the usual Friedmann equation $\dot{H}-k/a^{2}=-4\pi G(\rho+p)$.

Instead of solving exactly this differential equation to obtain $H(t)$, we shall examine, as it was done in Ref.~\cite{Awad}, the behavior of $\dot{H}$ by plotting the latter vs. $H$. For simplicity, we shall assume a spatially flat FLRW metric by taking $k=0$. Also, we shall consider the simple case of a radiation dominated Universe for which we have the equation of state $p=\rho/3$. With these assumptions, (\ref{14}) becomes
\begin{multline}\label{15}
\dot{H}=\left[\frac{4\pi}{\beta H^{2}}+\frac{\eta\sqrt{4\pi}}{|H|}
-\sqrt{\left(\frac{4\pi}{\beta H^{2}}+\frac{\eta\sqrt{4\pi}}{|H|}\right)^{2}-\frac{4\pi}{\beta H^{2}}}\right]\times
\\\Biggl(-2H^{2}-\frac{2\beta\eta|H|^{3}}{3\sqrt{\pi}}+\frac{2\xi^{3/2}|H|^{3}}{3\beta(1-\beta\eta^2)\sqrt{\pi}}
+\frac{2\eta[(\beta\eta^{2}-1)H^{2}+2\eta\sqrt{\pi}|H|]\xi^{1/2}}{(1-\beta\eta^{2})^{2}}
\\+\frac{8\pi\eta}{(1-\beta\eta^{2})^{5/2}}\sin^{-1}\left[\frac{\sqrt{\beta}(\beta\eta^{2}-1)|H|}{2\sqrt{\pi}}+\eta\sqrt{\beta}\right]-8\pi C\Biggr),
\end{multline}
with the constant of integration $C$ as given by (\ref{9}), and
\begin{equation}\label{16}
\xi=\frac{4\pi}{H^{2}}+\frac{4\beta\eta\sqrt{\pi}}{|H|}+\beta^{2}\eta^{2}-\beta.
\end{equation}

\begin{figure}[H]
\centering\includegraphics[angle=0, scale=0.9]{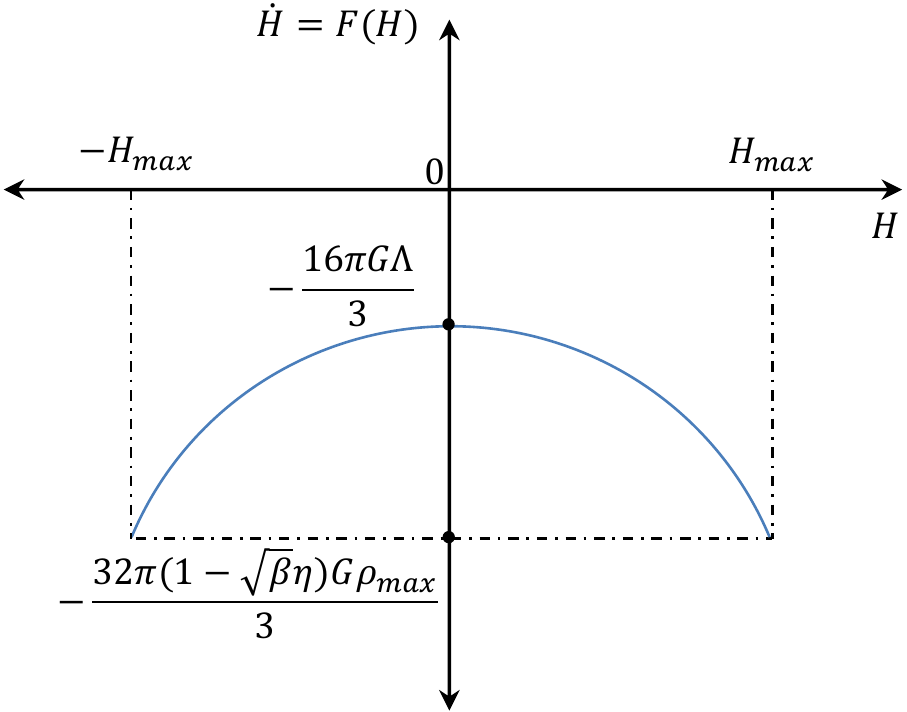}\\
\caption{Plot showing the shape of $\dot{H}=F(H)$.}\label{Fig1}
\end{figure}

The modified Friedmann equation (\ref{15}) does not look simple and it is therefore important to see explicitly what correction to the usual Friedmann equation, $\dot{H}=-2H^2$, obtained within general relativity for a radiation dominated epoch, does equation (\ref{15}) bring during the late-times expansion of the Universe, \textit{i.e.}, when $H\ll H_{max}\sim l_P^{-1}$. In fact\footnote{As rightly pointed out by the anonymous referee.}, it is hard to tell if the specific cyclic evolution, which we are going to argue for below Fig.~\ref{Fig1}, could be uniquely connected with this modified GUP or rather explained differently. For such a question, one needs\footnote{As suggested by the referee.} to study cosmological perturbations in detail and compare with COBE and Planck data. Therefore, we think that finding the first corrections to GR for small $H$\footnote{As requested by the referee.}, will certainly constitute the first step towards such a goal.

For $H$ such that $H\ll l_P^{-1}$, we can perform a Taylor expansion of (\ref{15}) around small values of $H$. The expansion can be found either from (\ref{15}) or, more simply, by expanding (\ref{f'(A)}) in terms of the apparent area $A$ around $A\rightarrow\infty$ and then substituting in (\ref{FR2}) and (\ref{FR1}). The result is,
\begin{equation}\label{FriedmannApprox}
\dot{H}=-2H^2+\frac{\beta\eta}{\sqrt{\pi}}|H|^3-\frac{\beta(8\beta\eta^2+3)}{48\pi}H^4+\mathcal{O}(H^5)-\frac{16\pi G\Lambda}{3}.
\end{equation}
We see that the correction terms to the familiar Friedmann equation are proportional to the constants $\beta$ and $\eta$ and, hence, to the parameters $\alpha$, $\mu$ and the Planck length $l_{P}$ rising from the modified GUP (\ref{1}). When we let the Planck length and/or the Hubble parameter tend to zero, we recover the usual Friedmann equation but with an additional term which is responsible here for the contracting phase of the Universe.

In fact, an important consequence of this result is that while the cosmological constant $\Lambda$ is positive and very small, as usual, to be in agreement with observations, the fact that it appears with a minus sign here is what causes the departure of the modified Friedmann equation from the usual one obtained within general relativity. Indeed, when adding a positive cosmological constant $\Lambda$ to a flat Universe filled with radiation the usual Friedmann equations obtained within general relativity are $H^2=8\pi G(\rho+\Lambda)/3$ and $\dot{H}=-2H^2+16\pi G\Lambda/3$. This implies that for a diluted enough radiation, i.e. when $\rho\rightarrow0$, the expansion of the Universe reaches a steady state fixed by the value of $\Lambda$. In our case, however, no such steady state will ever be reached since we end up with $H\rightarrow0$, and therefore $\dot{H}\rightarrow-16\pi G\Lambda/3$ according to (\ref{FriedmannApprox}). Hence, a contracting phase ensues instead. It is therefore the positivity of the cosmological constant $\Lambda$ itself that is responsible for the appearance of the contracting phase. The full evolution of the Universe through the various cycles is examined below.

Below, we have plotted the function $\dot{H}=F(H)$ in terms of the Hubble rate $H$, obtained from the full expression on the right-hand side of (\ref{15}).

From this graph, we see that the rate of change of $H$ stays negative within the whole range of allowed values for the Hubble parameter, \textit{i.e.}, from $H_{max}$ and all the way to $-H_{max}$. Starting from the right at the maximum value $H_{max}$, the rate of change $\dot{H}$ decreases in absolute value but stays nevertheless negative. This simply means that the Hubble parameter keeps decreasing until it completely vanishes. The corresponding value then for $\dot{H}$ is $-16\pi G\Lambda/3$. Thus, when the Hubble parameter vanishes we find that, because its rate of change is still negative, the Hubble flow should change sign and become negative too. This just means that the Universe enters a new phase which is a contracting phase.

Next, following the curve towards the left, we see that since the velocity is negative and increases in absolute value, the Hubble flow keeps becoming more negative faster, \textit{i.e.}, the contraction accelerates, until the maximum allowed density $\rho_{max}$ is reached again.
Now, at that point of the graph the following issue arises. The velocity $\dot{H}$ is still negative whereas the Hubble parameter is not allowed to decrease further given that it has reached the minimum value, $-H_{max}$, it is allowed to take on. Normally, however, in order for the Hubble parameter to reach such a minimum (or the maximum $H_{max}$ on the other side of the graph) one expects the rate of change $\dot{H}$ to vanish at those points of the graph. The only way to solve this issue, and still rely on our equations above, is to go back and look closer at the primary cause behind such peculiar behavior in the dynamics of the Universe.

In fact, the origin of the negative sign of $\dot{H}$ in the graph came from Eq.~(\ref{FR1}), for the left-hand side of that equation is negative whereas the factor $f'(A)$ is strictly positive. That equation, in turn, was found from the first law $\delta Q=\mathrm{d}E-W\mathrm{d}V$, where we used $\delta Q=T\mathrm{d}S$. Examining the variation of entropy $\mathrm{d}S$ in the latter identity closer, however, reveals that if taken as it is without carefully paying attention to the signs, an issue with the second law of thermodynamics arises. Indeed, we clearly see from Eq.~(\ref{dS}) that whenever $\mathrm{d}r_{A}$ is negative, entropy decreases. In fact, to be more precise, we actually find that, after substituting $\mathrm{d}r_{A}=-Hr_{A}^{3}(\dot{H}-k/a^{2})\mathrm{d}t$ in Eq. (\ref{dS}), the latter is written as,
\begin{equation}\label{FirstdS}
\mathrm{d}S=-\frac{f'(A)}{G}Hr_{A}^{4}\left(\dot{H}-\frac{k}{a^{2}}\right)\mathrm{d}t.
\end{equation}
It is therefore clear that whenever both $-H$ and the content of the parenthesis are of different signs, as it is the case on the second branch of the graph in Fig.~\ref{Fig1}, we will have a decreasing entropy with time, which is in contradiction with the second principle of thermodynamics. In order therefore to remain in accordance with the second principle for all values of $H$, we shall define the variation of entropy for $H$ negative as the absolute value of the right-hand side of Eq.~(\ref{FirstdS}). Therefore, whenever $H$ is negative, as it is the case on the left branch of Fig.~\ref{Fig1}, the variation of entropy should be written more correctly as
\begin{equation}\label{NewdS}
\mathrm{d}S=-\frac{f'(A)}{G}Hr_{A}^{4}\Big|\dot{H}-\frac{k}{a^{2}}\Big|\mathrm{d}t.
\end{equation}

Doing so, however, will entail to choose for the second principle of thermodynamics the following sign convention: $\delta Q=-T\mathrm{d}S$. This simply amounts to rely again on the second principle of thermodynamics but taking into account the fact that the Universe is in a contracting phase, for as we saw the identity worked perfectly for the expanding Universe but for the contracting Universe an issue in the dynamics arose. Physically, this could be understood by recalling that during the contracting phase the Universe should be gaining heat. To obtain that from a negative $T$ but a positive $\mathrm{d}S$ one in fact only needs to write the equality with a different sign convention.

Now, when the form (\ref{NewdS}) for entropy variation and the present sign choice for $\delta Q$ are substituted in the first law $\delta Q=\mathrm{d}E-W\mathrm{d}V$, the latter reads,
\begin{equation}\label{NewdotH}
\Big|\dot{H}-\frac{k}{a^{2}}\Big|f'(A)=4\pi G(\rho+p).
\end{equation}

Notice that this latter form is still consistent with the fact that for negative values of $H$, the rate of change $\dot{H}$ on the second branch of Fig.~\ref{Fig1} comes out negative too. It is just that the equation is capable of giving us the rate of change only up to a sign, which, in this case, should be deduced from the one it had in the previous phase.

This simple sign ambiguity actually makes all the difference in the resulting dynamics. Indeed, the fact that the dynamical equation (\ref{NewdotH}) contains information on the dynamics of the Hubble parameter only up to a sign means that the full dynamics could not be obtained from thermodynamics alone. This observation allows us in fact to solve the issue we discussed above, concerning the fate of the Hubble parameter, when the latter reaches its minimum value $-H_{max}$. Indeed, we may extrapolate Eq.~(\ref{NewdotH}) beyond that point just by removing the absolute value symbol there without violating in any case the equation itself. At $H=-H_{max}$, therefore, we will have the velocity $\dot{H}$ in Fig.~\ref{Fig1} acquire the inverse sign as it sweeps the same values of the Hubble parameter towards the right. The resulting new curve will then be the symmetric, with respect to the $H$-axis, of the previous one and appear as another branch in the plot. The full plot is depicted in the figure below.
\begin{figure}[H]
\centering\includegraphics[angle=0, scale=0.9]{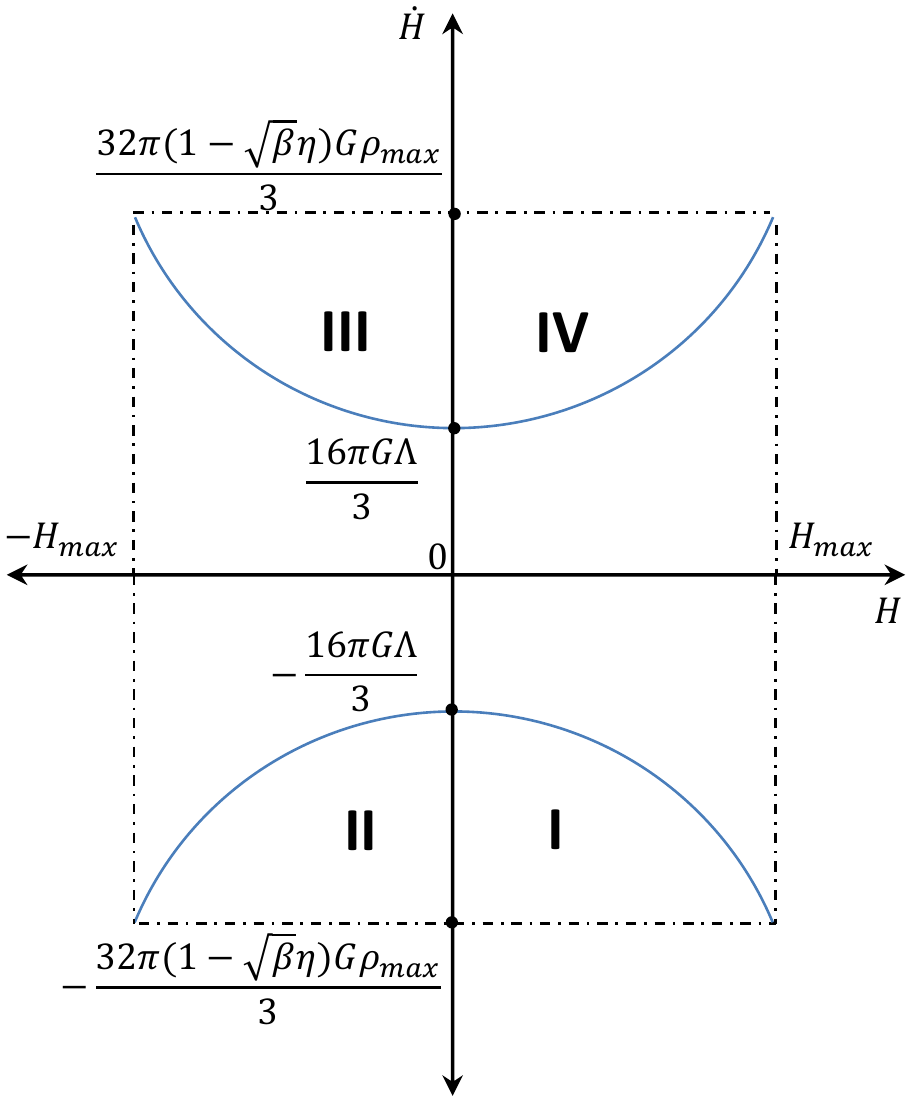}\\
\caption{The shape of $\dot{H}(H)$ after extrapolating the equations beyond the value $H=-H_{max}$.}\label{Fig2}
\end{figure}

Let us now provide the interpretation of the whole graph in Fig.~\ref{Fig2} with its four branches starting from the bottom-right. The Universe starts with an initial maximum density $\rho_{max}$ and a positive Hubble parameter $H_{max}$. It expands along the first branch while its rate of expansion $H$ keeps decreasing until it vanishes at the top of the lower curve. There, the Universe stops from expanding and starts contracting. The rate of contraction keeps increasing along the second branch as the absolute value $H$ keeps increasing too. The Universe then reaches its maximum rate of contraction $-H_{max}$ but not yet its maximum density $\rho_{max}$.

Indeed, thanks to the presence of the multivalued inverse sine in identity (\ref{7}), the values that the density $\rho$ will acquire when the apparent horizon's area $A$ starts decreasing, \textit{i.e.}, when the Universe starts contracting, might be different from those acquired during the expansion phase when passing over the same absolute values of the Hubble parameter $H$. Using this fact will actually make the reverse evolution consistent without preventing the density $\rho$ from being a continuous function. This is possible thanks to the fact that the constant of integration $C$ in (\ref{9}) is bound to respect the same choice for the inverse function $\sin^{-1}$ as the one appearing in the density $\rho$. Given that one can always make the redefinition $\sin^{-1}(x)\rightarrow-\sin^{-1}(x)+(2m+1)\pi$ for any integer $m$, this implies that the density in (\ref{7}) could be found only up to $-(2m+1)\pi\eta/[\sqrt{\beta}(1-\beta\eta^{2})^{5/2}]$ but will nevertheless be continuous when it reaches its minimum value $\rho_{min}$ for $A\rightarrow\infty$, as long as the constant of integration $C$ in (\ref{9}) follows also this redefinition of the $\sin^{-1}$ appearing there. Thus, when $\rho$ reaches the value $\rho_{min}$, the Universe starts contracting but the subsequent evolution of the density will be governed by the following equation,
\begin{eqnarray}
\frac{4G\rho}{3}&=&\frac{1}{A}+\frac{2\beta\eta}{3A^{3/2}}-\frac{2\xi^{3/2}}{3\beta(1-\beta\eta^2)A^{3/2}}
-\frac{\eta(\beta\eta^{2}-1+\eta\sqrt{A})\xi^{1/2}}{(1-\beta\eta^{2})^{2}A}\nonumber
\\&&+\frac{\eta}{\sqrt{\beta}(1-\beta\eta^{2})^{5/2}}\sin^{-1}\left[\frac{\sqrt{\beta}(\beta\eta^{2}-1)}{\sqrt{A}}+\eta\sqrt{\beta}\right]+C,
\end{eqnarray}
Therefore, the density $\rho$ will indeed not acquire exactly the same values it had previously during the expansion phase when passing again through the same absolute values of $H$ since it always acquires smaller values than during the previous phase.

Next, after reaching the lowest point of the second branch, as we saw, the subsequent dynamical evolution of the Hubble rate will be dictated by the third branch in Fig.~\ref{Fig2}. Thus, as the velocity $\dot{H}$ becomes now positive on this branch, the Hubble parameter increases from $-H_{max}$ to zero. During this phase, however, the Universe keeps contracting, since the Hubble flow remains negative during all the phase and the Universe has not yet reached its maximum density. The Hubble flow remains negative but changes in absolute value. This means that the contraction will still be taking place, only less quickly. It is only when the Hubble parameter vanishes at the end of the third branch, that the Universe stops contracting and the Big Crunch comes to an end.

Since the rate of change $\dot{H}$ at the end of the third branch is still positive, the Hubble parameter rises again from zero and starts increasing towards the right. This is where the bounce happens. This forms the fourth branch of the graph. During this phase the Universe expands from its much contracted state and slowly increases in size, but that happens faster and faster as time goes by because the Hubble parameter keeps increasing too. When the highest point of the fourth branch in the upper-right is reached, the Universe achieves its maximum expansion rate but with a different maximum density $\rho_{max}$, due again to its dependence on the multivalued inverse sine. At that point, the velocity flips sign again so that the remaining phase will be described by the first branch, starting from the bottom-right, and the process described above will recommence again.

Here we would like to emphasize the fact that, although the curve is traversed reversibly from negative to positive values of $H$, it should be kept in mind that this just shows the evolution of the Hubble flow. So at the end of the fourth branch, the size of the Universe is still very tiny so that everything looks as if a Big Bang is taking place again, exhibiting a sudden accelerated expansion, when in fact the expansion started way before; it just reached its maximum rate there.

Now, this looks pretty much like the familiar cyclic Universe scenario (see e.g. \cite{ashtekar, Science}), where here the Big Bang and Big Crunch happen many times but with different initial and final densities each time. It is important to keep in mind though that although this conclusion has been reached using the first and second law of thermodynamics, the details of the dynamics, \textit{i.e.}, the origin of the phase transitions responsible for the discontinuity in the rate of change $\dot{H}$, could only be inferred but not found from thermodynamics alone.

\section{Conclusion}
In this paper, we have applied the new version of the generalized uncertainty principle (GUP), recently introduced in the literature, to the study of the dynamics of the Universe. In contrast to the standard GUP which generalizes Heisenberg's uncertainty principle by adding to the right-hand side of the inequality a term quadratic in the uncertainty on momentum $\Delta p$, the version of the GUP we used in this paper contains one more additional term which is linear in the uncertainty $\Delta p$. As such, and given that the standard GUP has already been successfully applied in the literature to investigate its consequences for cosmology, it appears of considerable interest to examine what new consequences, if any, such a modification of GUP would bring for cosmology and what it could add to our current understanding of the latter.

This analysis was performed using Jacobson's approach. The modified GUP deformed the
    entropy-area relation    for the apparent horizon,
and this in turn produced correction terms for the Friedmann equations.
We were able to obtain, just as it was the case with the standard GUP, an upper bound for the density of the Universe,
using these modified Friedmann equations. Just as it was found in Ref.~\cite{Awad}, although with different numerical factors, the maximum density we found here (as it follows from Eqs.~(\ref{9}) and (\ref{12}), after keeping only the leading terms when $l_{P}\rightarrow0$) is also of the order of $l^{-2}_{P}$. This was used as an argument for the absence of any singularity at the Big Bang. Now it is well-known that the existence of a maximum density for the Universe already emerges in loop quantum cosmology within which one finds for $\rho_{max}$ a value of the order $\sim0.41\rho_{_{P}}$, where $\rho_{_{P}}$ is the Planck density \cite{ashtekar}. In our case, the specific value we find from Eqs.~(\ref{12}) and (\ref{9}), after keeping only the leading terms, is $\rho_{max}=5\rho_{_{P}}/[8\pi\alpha^{2}(1+\mu)]$. In Ref.~\cite{Awad}, however, the result was $\rho_{max}=5\rho_{_{P}}/(4\pi\beta)$, where the dimensionless parameter $\beta$ was left unspecified there. This difference might actually be of great importance when one tries to decide between the two versions of the GUP to use to study the physics of the early Universe, and even for confronting our results with those of loop quantum cosmology to establish the correctness of our approach. A rigorous analysis of this point will be attempted in a separate work. Nevertheless, it is already clear that the present method reproduces the cyclic Universe scenario arising both in loop quantum cosmology \cite{ashtekar} and the Ekpyrotic scenario \cite{Science}, whereas it is impossible to consistently make it emerge within the purely quadratic version of the GUP due to the absence of the multivalued function $\sin^{-1}$ there \cite{Awad}.

Where the two approaches, the one based on the standard GUP and the present one based on the modified GUP, also differ is on the details of the dynamics for the FLRW Universe. The rate of change of the Hubble parameter implied by the modified GUP-based approach is slightly different from what is found when using the standard GUP (compare (\ref{15}) with Eq.~(6.3) of Ref.~\cite{Awad}). Such a difference, albeit very small, could be used as a means to distinguish experimentally between the two versions of the GUP, as used in the framework of our approach, since this result is particularly relevant to the early inflationary expansion of the Universe.

It is not hard to conduct a similar analysis within other cosmological models. It would be interesting for example to analyze the effect such a deformation of the Friedmann equations could have on Bianchi Cosmologies. Furthermore, it is expected that this approach will also deform the Wheeler-DeWitt equation for the Universe. It would therefore also be interesting to investigate the consequences of this approach on the Wheeler-DeWitt equation. Indeed, any modification of the latter is expected to bring non-trivial modifications to most models of quantum cosmology \cite{qc}. In fact, it would certainly be very illuminating to analyze first some simple models of quantum cosmology within the framework of the present approach.

Finally, we would like to recall here the fact that the modified GUP used in this paper is actually consistent with other axes of research investigating the physics of space-time, such as, discrete space-time \cite{17}, spontaneous symmetry breaking of Lorentz invariance in string field theory \cite{18}, ghost condensation \cite{18aa}, space-time foam models \cite{19}, spin-networks in loop quantum gravity \cite{20}, non-commutative geometry \cite{21,21aa}, and Horava-Lifshitz gravity \cite{22}. Therefore, another way to assess the correctness of the approach developed in this paper is to confront our results found here with similar case studies based on these other axes of research besides those of loop quantum cosmology discussed above. In addition, since phase-space non-commutativity has been introduced based on the generalized uncertainty principle \cite{Bina}, it might also be of interest to adapt our approach to other similar physical situations as a gravitational collapse of a star which has been studied recently in Ref.~\cite{Rasouli1} or to study cosmic expansion in alternative theories of gravity as done in Ref.~\cite{Rasouli2}, both based on non-commutativity in phase-space. Any sign of disagreement could then be used either, to rectify the new version of GUP adopted here, or to reexamine the whole approach followed here which consists in applying horizon thermodynamics to the Universe's apparent horizon.
\\\\
{\bf \large{Acknowledgments}}\\ \\ \noindent
We would like to thank the anonymous referee for his/her valuable comments that helped improve the clarity of this paper. The research of AFA is supported by Benha University (www.bu.edu.eg).


\begin{thebibliography}{}
\bibitem{Mead} C. A. Mead, Phys. Rev. \textbf{B849}, 135 (1964)
\bibitem{Veneziano} G. Veneziano, Europhys. Lett. 2, 199 (1986)
\bibitem{Amati} D. Amati, M. Ciafaloni and G. Veneziano, Phys. Lett. B \textbf{197}, 81 (1987)
\bibitem{Gross} D. J. Gross and P. F. Mende, Phys. Lett. B \textbf{197}, 129 (1987)
\bibitem{Yonega} T. Yonega, Mod. Phys. Lett. A \textbf{4}, 1587 (1989)
\bibitem{Konishi} K. Konishi, G. Paffuti and P. Provero, Phys. Lett. B \textbf{234}, 276 (1990)
\bibitem{Guida} R. Guida and K. Konishi, Mod. Phys. Lett. A \textbf{6}, 1487 (1991)
\bibitem{Maggiore1} M. Maggiore, M. Maggiore, Phys. Lett. \textbf{B304}, 65 (1993)
\bibitem{Maggiore2} Maggiore, Phys. Rev. \textbf{D49}, 5182 (1994)
\bibitem{Maggiore3} M. Maggiore, Phys. Lett. \textbf{B319}, 83 (1993)
\bibitem{Garay} L. J. Garay, Int. J. Mod. Phys. \textbf{A10}, 145 (1995)
\bibitem{Kempf1} A. Kempf, G. Mangano, R. B. Mann, Phys. Rev. \textbf{D52}, 1108 (1995)
\bibitem{Kempf2} A. Kempf, J. Phys. \textbf{A30}, 2093 (1997)
\bibitem{Brau} F. Brau, J. Phys. \textbf{A32}, 7691 (1999)
\bibitem{Scardigli} F. Scardigli, Phys. Lett. \textbf{B452}, 39 (1999)
\bibitem{Hossenfelder1} S. Hossenfelder, M. Bleicher, S. Hofmann, J. Ruppert, S. Scherer and H. Stoecker, Phys. Lett. \textbf{B575}, 85 (2003)
\bibitem{Bambi} C. Bambi and F. R. Urban, Class. Quant. Grav. \textbf{25}, 095006 (2008)
\bibitem{Ali1} A. F. Ali, S. Das and E. C. Vagenas, Phys. Lett. \textbf{B678}, 497 (2009)

\bibitem{Ali:2011ap}
  A.~F.~Ali,
  Class.\ Quant.\ Grav.\  {\bf 28}, 065013 (2011)

\bibitem{Ali:2011fa}
  A.~F.~Ali, S.~Das and E.~C.~Vagenas,
  Phys.\ Rev.\ D {\bf 84}, 044013 (2011)

\bibitem{Das:2010zf}
  S.~Das, E.~C.~Vagenas and A.~F.~Ali,
  Phys.\ Lett.\ B {\bf 690}, 407 (2010);
  Phys.\ Lett.\  {\bf 692}, 342 (2010)

\bibitem{Ali:2012mt}
  A.~F.~Ali,
  JHEP {\bf 1209}, 067 (2012)
\bibitem{Hossenfelder}   S.~Hossenfelder,
  Living Rev.\ Rel.\  {\bf 16}, 2 (2013)
\bibitem{Medved} A. J. M. Medved and E. C. Vagenas, Phys. Rev. D \textbf{70}, 124021 (2004)
\bibitem{Ali2} A. F. Ali and A. Tawfik, Adv. High Energy Phys. 126528 (2013)
\bibitem{Awad} A. Awad and A. F. Ali, JHEP \textbf{1406}, 093 (2014)
\bibitem{Bina} A. Bina, S. Jalalzadeh and A. Moslehi, Phys. Rev. D \textbf{81}, 023528 (2010)

\bibitem{Akbar} M. Akbar and R. -G. Cai, Phys. Rev. \textbf{D75}, 084003 (2007)
\bibitem{Majumder} B. Majumder, Astrophys. Space Sci. \textbf{336}, 331 (2011)
\bibitem{Lidsey} J. E. Lidsey, Phys. Rev. \textbf{D88}, 103519 (2013)
\bibitem{Jalalzadeh} S. Jalalzadeh, S. M. M. Rasouli and P. V. Moniz, Phys. Rev. D \textbf{90}, 023541 (2014)
\bibitem{DSR}J. Magueijo, and L. Smolin, Phys. Rev. Lett. \textbf{88}, 190403 (2002)
\bibitem{DSR1} J. Magueijo, and L. Smolin, Phys. Rev. D \textbf{71}, 026010 (2005)
\bibitem{17} G. 't Hooft, Class. Quant. Grav.  \textbf{13}, 1023 (1996)
\bibitem{18}V. A. Kostelecky and S. Samuel, Phys. Rev. D  \textbf{39}, 683 (1989)
\bibitem{18aa}M. Faizal, J. Phys. A  \textbf{44}, 402001 (2011)
\bibitem{19} G. Amelino-Camelia, J. R. Ellis, N. E. Mavromatos, D. V. Nanopoulos and S. Sarkar, Nature  \textbf{393},
763 (1998)
\bibitem{20}R. Gambini and J. Pullin, Phys. Rev. D  \textbf{59}, 124021 (1999)
\bibitem{21} S. M. Carroll, J. A. Harvey, V. A. Kostelecky, C. D. Lane and T. Okamoto, Phys. Rev. Lett.  \textbf{87},
141601 (2001)
\bibitem{21aa}M. Faizal,  Mod. Phys. Lett. A  \textbf{27}, 1250075 (2012)
\bibitem{22} P. Horava, Phys. Rev. D  \textbf{79}, 084008 (2009)
\bibitem{cai} R.-G. Cai and S.P. Kim, JHEP \textbf{02}, 050 (2005)

\bibitem{Lima} J. A. S. Lima and S. H. Pereira, Phys. Rev. D \textbf{78}, 083504 (2008)

\bibitem{Hayward:1997jp}
  S.~A.~Hayward,
   Class.\ Quant.\ Grav.\  {\bf 15}, 3147 (1998)



\bibitem{Bak} D. Bak and S-J. Rey, Class. Quant. Grav. \textbf{17}, L83 (2000)
\bibitem{Viaggiu} S. Viaggiu, Gen. Rel. Gravit. \textbf{47} 8 (2015)
\bibitem{tHooft} G. 't Hooft, Dimensional Reduction in Quantum Gravity, in "Salamfest" pp. 284-296 (World
Scientific Co, Singapore, 1993).
\bibitem{Susskind} L. Susskind, J. Math. Phys. 36, 6377 (1995)
\bibitem{Fischler} W. Fischler and L. Susskind, Holography and Cosmology, 
\bibitem{Easson1} D. A. Easson, P. H. Frampton and G. F. Smoot, Phys. Lett. B \textbf{696}, 273
(2011)
\bibitem{Easson2} D. A. Easson, P. H. Frampton and G. F. Smoot, Int. J. Mod. Phys. A \textbf{27},
1250066 (2012)
\bibitem{Komatsu} N. Komatsu and S. Kimura, Phys. Rev. D 88, 083534 (2013)
\bibitem{hammad} F. Hammad, Class. Quantum Grav. \textbf{30} 125011 (2013)

\bibitem{Wang} B. Wang, Y. Gong and E. Abdalla, Phys.Rev. D \textbf{74}, 083520 (2006)

\bibitem{Cai2} R-G. Cai, L-M. Cao, Y-P. Hu, Class. Quantum Grav. \textbf{26}, 155018 (2009)
\bibitem{Parikh} M. K. Parikh and F. Wilczek, Phys. Rev. Lett. \textbf{85}, 5042 (2000)

\bibitem{Jacobson1} T. Jacobson, Phys. Rev. Lett. \textbf{75}, 1260 (1995).

\bibitem{Carloni} S. Carloni, R. Goswami and P. K. S. Dunsby, Class. Quantum Grav. \textbf{29}, 135012 (2012)


\bibitem{adler} R. J. Adler, P. Chen, D. I. Santiago, Gen. Rel. Grav. {\bf 33}, 2101-2108 (2001). 



\bibitem{Ali:2013qza}
  A.~F.~Ali,
  Phys.\ Lett.\ B {\bf 732}, 335 (2014)

\bibitem{Camelia}G. Amelino-Camelia, M. Arzano and A. Procaccini, Phys. Rev. {\bf D 70}, 107501 (2004); E.M. Lifshitz, L.P. Pitaevskii and V.B. Berestetskii, LandauLifshitz, Course of Theoretical Physics, Volume 4: Quantum Electrodynamics, (Reed Educational and Professional Publishing, 1982).

\bibitem{MDR} J. Alfaro, H.A. Morales-Tecotl, L.F. Urrutia, Phys. Rev. Lett. \textbf{84} (2000) 2318; G. Amelino-Camelia, M. Arzano, A. Procaccini, Int. J. Mod. Phys. D \textbf{13} (2004) 2337;
\bibitem{Amelino} G. Amelino-Camelia, M. Arzano, Y. Ling and G. Mandanici, Class. Quant. Grav. \textbf{23}, 2585 (2006)
\bibitem{Xiang} Li Xiang, Phys. Lett. B \textbf{638}, 519 (2006)

\bibitem{ashtekar} A. Ashtekar and P. Singh, Class. Quant. Grav. \textbf{28}, 213001 (2011)
\bibitem{qc} M.  Bojowald, Rep. Prog. Phys. 78, \textbf{023901} (2015)
\bibitem{Science} P. J. Steinhardt and N. Turok, Science \textbf{296}, 1436 (2002)
\bibitem{Rasouli1} S. M. M. Rasouli, A. H. Ziaie, J. Marto and P. V. Moniz, Phys. Rev. D \textbf{89}, 044028 (2014)
\bibitem{Rasouli2} S. M. M. Rasouli and P. V. Moniz, Phys. Rev. D \textbf{90}, 083533 (2014)

\end{thebibliography}
\end{document}